\documentclass[twocolumn]{aastex631}

\usepackage{amsmath}
\usepackage{hyperref}
\usepackage{booktabs}    
\usepackage{threeparttable}

\shortauthors{Liu et al.}

\begin{document}

\title{Probing the $\gamma$-ray Emission Origin of Two Star-forming Galaxies NGC 2403 and NGC 3424 with the Fermi-LAT}

\author[0009-0004-6750-821X]{Linjie Liu}
\affiliation{Yunnan Observatories, Chinese Academy of Sciences, Kunming 650216, China}
\affiliation{University of the Chinese Academy of Sciences, Beijing 100049, China}

\author[0000-0003-2839-1325]{Wei Zhang}
\affiliation{State Key Laboratory for Particle Astrophysics \& Experimental Physics Division \& Computing Center, Institute of High Energy Physics, Chinese Academy of Sciences, 100049 Beijing, China}
\affiliation{TIANFU Cosmic Ray Research Center, Chengdu, Sichuan, China}

\author[0000-0003-0933-6101]{Xian Hou}
\affiliation{Yunnan Observatories, Chinese Academy of Sciences, Kunming 650216, China}

\author[0000-0003-0933-6101]{Pierrick Martin}
\affiliation{IRAP, Universit\'e de Toulouse, CNRS, UPS, CNES, F-31028 Toulouse, France}

\email{xhou@ynao.ac.cn}
\email{pierrick.martin@irap.omp.eu}

\begin{abstract}
Star-forming galaxies (SFGs) are a subclass of $\gamma$-ray emitters and a correlation between their $\gamma$-ray luminosity ($L_{\rm \gamma}$) and the total infrared (IR) luminosity ($L_{\rm IR}$) has been established based on the \textit{Fermi} Large Area Telescope (LAT) data. 
NGC 2403 and NGC 3424 have been reported as outliers in the $L_{\rm \gamma}$-$L_{\rm IR}$ correlation with light curves showing significant variability, which contrasts with the temporally stable $\gamma$-ray emission in other SFGs, originating primarily from cosmic rays interacting with interstellar medium.
In this study, we reanalyze the $\gamma$-ray emission in the directions of NGC 2403 and NGC 3424 using more than 16.5 yr \emph{Fermi}-LAT data. 
NGC 3424 is found to be spatially coincident with the detected $\gamma$-ray source, while NGC 2403 is significantly offset from the nearest $\gamma$-ray source, suggesting an implausible association.
We confirm the previously reported variability of both $\gamma$-ray sources and the significant deviation from the $L_{\rm \gamma}$-$L_{\rm IR}$ correlation when assuming an association of both $\gamma$-ray sources with the two galaxies.
Our findings lend further support to the interpretation that their $\gamma$-ray emission is driven primarily by alternative radiative processes—rather than by star formation activity—such as the ejecta of the Type IIP supernova SN 2004dj in NGC 2403 interacting with a surrounding high-density shell and an obscured active galactic nucleus in NGC 3424.
\end{abstract}


\section{Introduction}
\label{introduction}

Star-forming galaxies (SFGs) are huge reservoirs of cosmic rays (CRs) that produce $\gamma$-rays primarily through interactions of CRs with the interstellar medium (ISM). Based on the data accumulated by the \textit{Fermi} Large Area Telescope \citep[LAT,][]{Atwood2009}, several SFGs have been detected \citep[See e.g., Table 1 in][]{Ajello+etal+2020,Xiang+etal+2023,Xing+etal+2023}, and their $\gamma$-rays are found to be non-linearly correlated with their total infrared (IR) luminosity at $8-1000\, \rm \mu m$, an indicator of star formation rate \citep[SFR,][]{Kennicutt1998}. This correlation and the lack of $\gamma$-ray variability favor the idea that the $\gamma$-ray emission from SFGs is produced mainly by CR-ISM interactions.

However, there are complicating factors for the origin of the $\gamma$-ray emission of SFGs. For spatially unresolved SFGs (the majority of the detected sample, detected as a point-like source), the observed emission will arise from CR-ISM interactions, and also from unresolved (below the LAT detection threshold or confused with the diffuse emission of the host galaxy) populations of sources like pulsars and supernova (SN) remnants, with a far more speculative contribution from dark matter. Conversely, the Large Magellanic Cloud \citep[LMC,][]{Abdo2010LMC,Ackermann2015} and the Small Magellanic Cloud \citep[SMC,][]{Abdo2010SMC}, and to a lesser extent M31 \citep{Abdo2010M31,Ackermann+etal+2017,Xing+etal+2023}, are spatially resolved, and they allow for some separation of interstellar emission and discrete sources, at least the strongest sources, like the $\gamma$-ray pulsar detected in LMC \citep{Ackermann2015}.

NGC 2403 \citep[3.2 Mpc;][]{Tully2013} and NGC 3424 \citep[27.2 Mpc;][]{Theureau2007} are two SFGs that have been reported to show significant variability (2.7$\sigma$ and 4.1$\sigma$, respectively) in their long-term light curves using 11.4 yr \citep{Xi+etal+2020a} and 10.5 yr \citep{Peng+etal+2019} \emph{Fermi}-LAT data, respectively, contrary to the steady emission expected from CR-ISM process. \cite{Xi+etal+2020a} claimed that the $\gamma$-ray emission attributed to NGC 2403 may originate from the explosion of the near and bright Type IIP supernova SN 2004dj in NGC 2403 \citep{Nakano+etal+2004,Patat2004} which is spatially coincident with the detected gamma-ray source. 
\cite{Peng+etal+2019} proposed that NGC 3424 may harbor an active galactic nucleus (AGN) that is heavily obscured by the dust lanes of the host galaxy and the circumnuclear torus, and the AGN activity is likely the dominant contributor to its gamma-ray emission compared with the contribution from starburst activity.
Furthermore, both galaxies have been identified as outliers in the empirical $L_{\rm \gamma}$-$L_{\rm IR}$ correlation, lying above the calorimetric limit\footnote{A galaxy acts as a CR calorimeter when the majority of the energy carried by CRs is dissipates within the galaxy itself \citep{Loeb2006,Thompson2007,Lacki2011} } by factors of approximately 2 and 10, respectively \citep{Ajello+etal+2020,Xi+etal+2020a,Peng+etal+2019}. Together with the observed variability of the two galaxies, these findings suggest that their observed $\gamma$‑ray emission likely originates from additional radiation processes beyond those driven by the star‑formation activity.

Motivated by the uniqueness and the enigmatic origin of the $\gamma$-ray emission observed from NGC 2403 and NGC 3424, we revisited the $\gamma$-ray emission toward the two galaxies using more than 16.5 yr \emph{Fermi}-LAT data$-$approximately five to six more years data than employed in earlier studies$-$together with the most recent source catalog, which offers an improved and more accurate modeling of the $\gamma$‑ray sky. The aim of this work is to perform a thorough sanity check of the previously reported results through a refined analysis and ultimately resolve the puzzle concerning the origin of the $\gamma$‑ray emission from these two galaxies. The paper is organized as follows. In Section ~\ref{Obs}, we briefly introduce the dataset and the analysis methods. In Section ~\ref{data}, we present the spatial and spectral analysis results as well as the flux variability. We then discuss our results in Section ~\ref{discussion} and finally conclude in Section ~\ref{conclusion}.

\section{Dataset And Analysis Methods}
\label{Obs}

We used more than 16.5 yr \emph{Fermi}-LAT Pass 8 data \citep{Atwood2013,Bruel2018} collected from 2008 August 4 to 2025 March 5 (MET 239557417 - MET 762825605) in the 0.1$–$800 GeV energy range. SOURCE class events with good quality were selected by applying the standard event filter $\rm {(DATA\_QUAL > 0)\&\&(LAT\_CONFIG==1)}$\footnote{\url{https://fermi.gsfc.nasa.gov/ssc/data/analysis/scitools/data\_preparation.html}}. Following the \textit{Fermi} Large Area Telescope Fourth Source Catalog \citep[4FGL;][]{4FGL}, we used specific zenith angle cuts depending on the point-spread function (PSF) and energy range to reduce the contamination from the low-energy Earth limb emission. We retained PSF 2 and PSF 3 event types with zenith angles \(\textless 90^\circ\) in the energy range of 0.1$-$0.3 GeV, PSF 1, PSF 2 and PSF 3 event types with zenith angles \(\textless 100^\circ\) in the energy range of 0.3$-$1 GeV, and all events with zenith angles \(\textless 105^\circ\) at above 1 GeV. We also excluded time intervals when Solar flares and Gamma-ray Bursts (GRBs) occurred \citep{4FGL-DR3,4FGL-DR4}.

We performed a binned summed maximum likelihood analysis using Fermitools\footnote{\url{https://fermi.gsfc.nasa.gov/ssc/data/analysis/software/}} (v2.2.0) and the Fermipy package\footnote{\url{https://fermipy.readthedocs.io/en/latest/index.html}} \citep[v1.3.1;][]{Wood+etal+2017}. The analysis was performed within a region of interest (ROI) of size $14^{\circ}\times14^{\circ}$ centered on the optical position of NGC 2403 (R.A. = $114\fdg21$, Decl. = $65\fdg60$) and the IR position of NGC 3424 (R.A. = $162\fdg94$, Decl. = $32\fdg90$) from SIMBAD\footnote{\url{https://simbad.u-strasbg.fr/simbad/}}, respectively. We binned the data with a pixel size of $0\fdg05$ and into ten energy bins per decade. 
For each ROI, sources within $20^{\circ}$ around NGC 2403 and NGC 3424 in the latest \emph{Fermi}-LAT source catalog 4FGL-DR4  \citep[gll\_psc\_v35.fit\footnote{\url{https://fermi.gsfc.nasa.gov/ssc/data/access/lat/14yr_catalog/}},][]{4FGL-DR3,4FGL-DR4} were included in the source model and we modeled the Galactic and isotropic diffuse emission\footnote{\url{https://fermi.gsfc.nasa.gov/ssc/data/access/lat/BackgroundModels.html}} using ``gll\_iem\_v07.fits'' and ``iso\_P8R3\_SOURCE\_V3\_v1.txt'', respectively. 
During the analysis, the spectral parameters of sources within $5^{\circ}$ of the ROI center were allowed to vary, as well as the normalizations of variable sources (those with a variability index exceeding 27.69, corresponding to a 99\% confidence level in the 4FGL-DR4 catalog) located between $5^{\circ}$ and $7^{\circ}$ from the ROI center. This choice was made in consideration of the broad \emph{Fermi}‑LAT PSF at 100 MeV as well as the potential influence of variable sources on the $\gamma$‑ray emission associated with our target galaxies. The parameters of all other sources were kept fixed at their catalog values. We also considered the energy dispersion effect, which becomes significant at low energies \citep{4FGL-DR3},  for all sources except for the isotropic component. 

We used the test statistic (TS) to characterize the detection significance of a source and claimed that a source is detected when it has a TS $>25$. The TS is defined as $2(\log \mathcal{L}_{1}-\log \mathcal{L}_{0})$, where $\log \mathcal{L}_{1}$ and $\log \mathcal{L}_{0}$ are the logarithms of the maximum likelihood of the complete source model and of the null-hypothesis model (i.e., the background model without the source included), respectively.
The TS$_{\rm ext}$, defined as $2(\log \mathcal{L}_{\rm e}-\log \mathcal{L}_{\rm p})$, where $\mathcal{L}_{\rm e}$ is the logarithm of the maximum likelihood of the fit with an extended spatial model and $\log \mathcal{L}_{\rm p}$ is that of the point-like model, was used to quantify the spatial extension of NGC 2403 and NGC 3424. A source is considered significantly extended if TS$_{\rm ext}>9$ in this work. 
The spectral curvature is quantified using TS$_{\rm curv} =2(\log \mathcal{L}_{\rm curv}-\log \mathcal{L}_{\rm PL})$, where $\mathcal{L}_{\rm curv}$ is the maximum likelihood of the curved spectral model and $\mathcal{L}_{\rm PL}$ is that of the power-law model. In this work, the spectrum is considered significantly curved if TS$_{\rm curv} >$ 9.
%
In the analysis for both galaxies, we first optimized each ROI (\emph{optimize} method in $\mathtt{fermipy}$), then searched for new $\gamma$-ray point sources and added to the source model when its TS $> 25$. Detailed spatial, spectral, and variability analyses have been performed for each galaxy.

\section{Data Analysis Results}
\label{data}

\subsection{Spatial Analysis}
\label{morphological}
The $\gamma$-ray spatial analysis was conducted in the 1$-$800 GeV range considering the better angular resolution and reduced background contamination.
The point source 4FGL J0737.4+6535 is located in the vicinity of NGC 2403 with an offset of $0\fdg07$, and 4FGL J1051.6+3253 is found near NGC 3424 with an offset of $0\fdg03$ \citep{4FGL-DR4}. 
To characterize the spatial consistency between the 4FGL $\gamma$-ray source and the target galaxy, we explored various geometrical models
for each 4FGL counterpart with its spectrum modeled by a simple power law (PL). 
First, we compared the point source models between fixed and relocalized positions of the 4FGL source (\emph{localize} method in $\mathtt{fermipy}$). 
This was followed by an extension test (\emph{extension} method in $\mathtt{fermipy}$) using a uniform disk and 2D Gaussian for both the fixed and relocalized point source models. 
Then, we tested the model where the $\gamma$-ray source position was fixed to the galaxy center. For NGC 2403, previous studies have revealed the existence of a Type IIP supernova SN 2004dj (R.A. = $114^{\circ}.321$, decl. = 65$^{\circ}.599$) in the 95\% confidence localization radius ($r_{95}$) of the $\gamma$-ray source \citep{Xi+etal+2020a}. Thus, we also tried to assess the spatial consistency of the $\gamma$-ray source with SN 2004dj by fixing the $\gamma$-ray source at the position of SN 2004dj.

The best-fit parameters of each model are summarized in Table~\ref{tab:Tab1} for NGC 2403 and Table~\ref{tab:Tab2} for NGC 3424, which include the negative log-likelihood ($-\log \mathcal{L}$) value of the model fit, the best-fit source position (R.A. and decl.), the $r_{95}$ of the best-fit position, the angular distance of each tested position from the galaxy, and the TS value of the $\gamma$-ray source. We found that for both galaxies, the relocalized point source model resulted in the smallest $-\log \mathcal{L}$ and highest TS, while the extension is not significant ($\mathrm{TS_{ext}} < 0.5$) using either the disk or Gaussian model, no matter which kind of position was used. 
Additionally, a two-component model consisting of a point source at the best-fit position and another point source fixed at the center of the galaxy was also tested. However, this model did not yield significant detection for the individual components (TS $\approx 10.5$ and $3.4$ in the case of NGC 2403; TS $\approx 1.6$ and $0.0$ in the case of NGC 3424), indicating that the additional component is not required.

\begin{deluxetable*}{lcccccc}[htbp]
\tabletypesize{\small} 
\tablewidth{\textwidth} 
\tablecaption{Spatial Analysis Results in 1--800 GeV for NGC 2403}
\label{tab:Tab1}
\tablehead{
\colhead{Spatial Model} & \colhead{$-\log \mathcal{L}$} & \colhead{R.A.} &
\colhead{decl.} & \colhead{$r_{\rm 95}$} & \colhead{$\theta$} & \colhead{TS} \\
& & \colhead{(deg)} & \colhead{(deg)} & \colhead{(deg)} & \colhead{(deg)} 
}
\startdata
Point Source (fixed 4FGL) & 269679.4 & 114.374 & 65.591 & \ldots & 0.068 & 46.8 \\
Point Source (free 4FGL)  & 269679.3 & 114.371 $\pm$ 0.023 & 65.601 $\pm$ 0.025 & 0.057 & 0.066 & 46.9 \\
Point Source (fixed NGC)  & 269682.8 & 114.21   & 65.60  & \ldots & 0 & 39.9 \\
Point Source (fixed SN)   & 269679.9 & 114.321  & 65.599 & \ldots & 0.046 & 45.7  \\
\enddata
\tablecomments{
The fixed 4FGL and free 4FGL models correspond to cases where the $\gamma$-ray source position was fixed to its 4FGL-DR4 catalog position and refitted (relocalized), respectively. The fixed NGC and fixed SN models correspond to cases where the $\gamma$-ray source position was fixed to the optical position of the galaxy and to that of SN~2004dj. The quantity $-\log \mathcal{L}$ denotes the negative log-likelihood value of the model fit. R.A. and Decl. are the best-fit source positions. $r_{\rm 95}$ is the 95\% confidence localization radius, and $\theta$ is the angular distance from the galaxy position. TS represents the detection significance of the $\gamma$-ray source.
}
\end{deluxetable*}

\begin{deluxetable*}{lcccccc}
\tabletypesize{\small}
\tablewidth{\textwidth}
\tablecaption{Spatial Analysis Results in 1--800 GeV for NGC 3424}
\label{tab:Tab2}
\tablehead{
\colhead{Spatial Model} & \colhead{$-\log \mathcal{L}$} & \colhead{R.A.} &
\colhead{Decl.} & \colhead{$r_{\rm 95}$} & \colhead{$\theta$} & \colhead{TS} \\
& & \colhead{(deg)} & \colhead{(deg)} & \colhead{(deg)} & \colhead{(deg)} 
}
\startdata
Point source (fixed 4FGL) & 159324.9 & 162.911 & 32.885 & \ldots & 0.028 & 28.9 \\
Point source (free 4FGL)  & 159324.8 & 162.903 $\pm$ 0.031 & 32.896 $\pm$ 0.028 & 0.072 & 0.031 & 29.1 \\
Point source (fixed NGC)  & 159325.5 & 162.94  & 32.90  & \ldots & 0 & 27.6 \\
\enddata
\tablecomments{
Same as Table~\ref{tab:Tab1}, but the fixed NGC here refers to the model in which the
$\gamma$-ray source position was fixed to the IR position of NGC 3424.
}
\end{deluxetable*}

\begin{figure*}[htbp]
\centering
\includegraphics[width=8.4cm, angle=0]{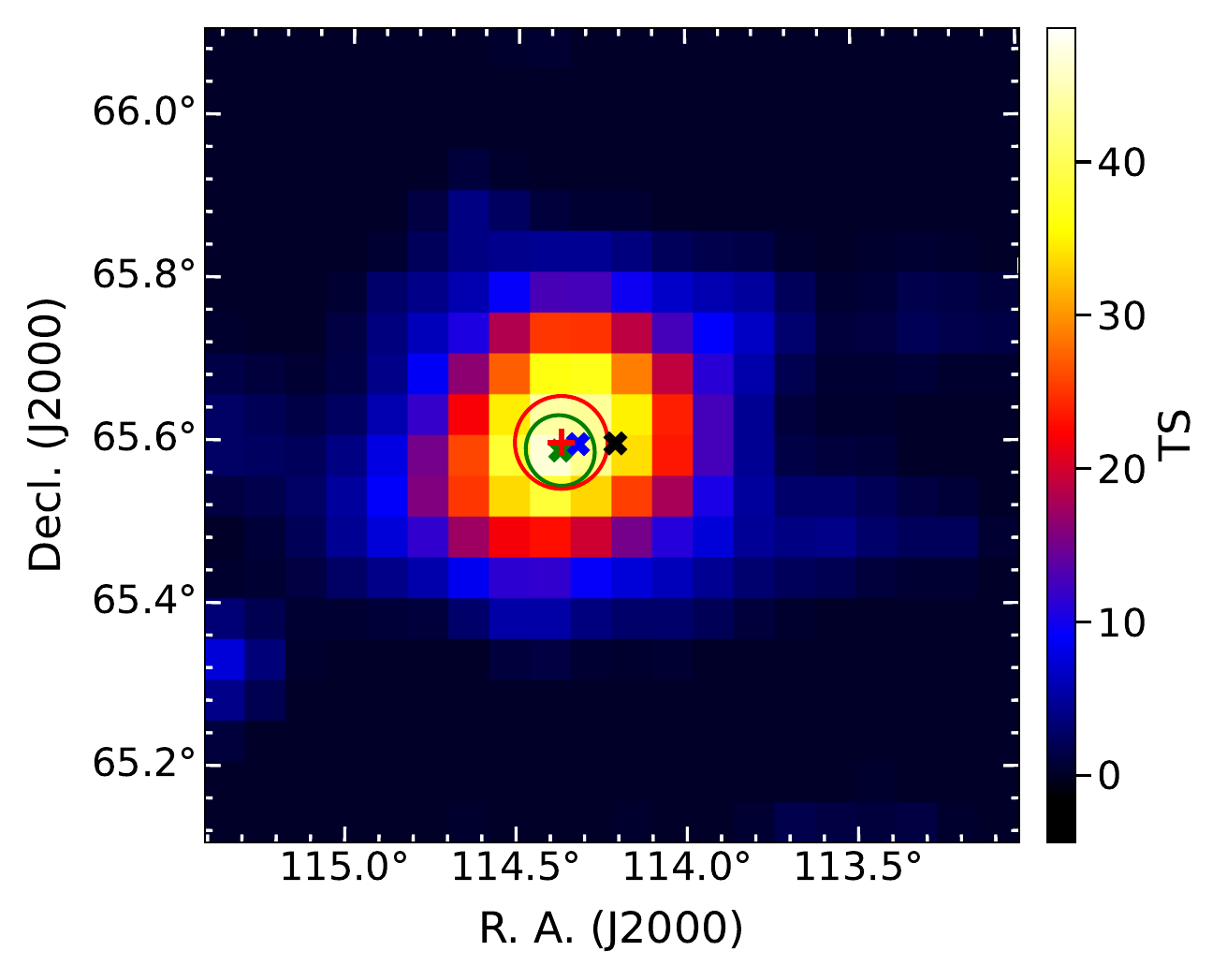}
\includegraphics[width=8.4cm, angle=0]{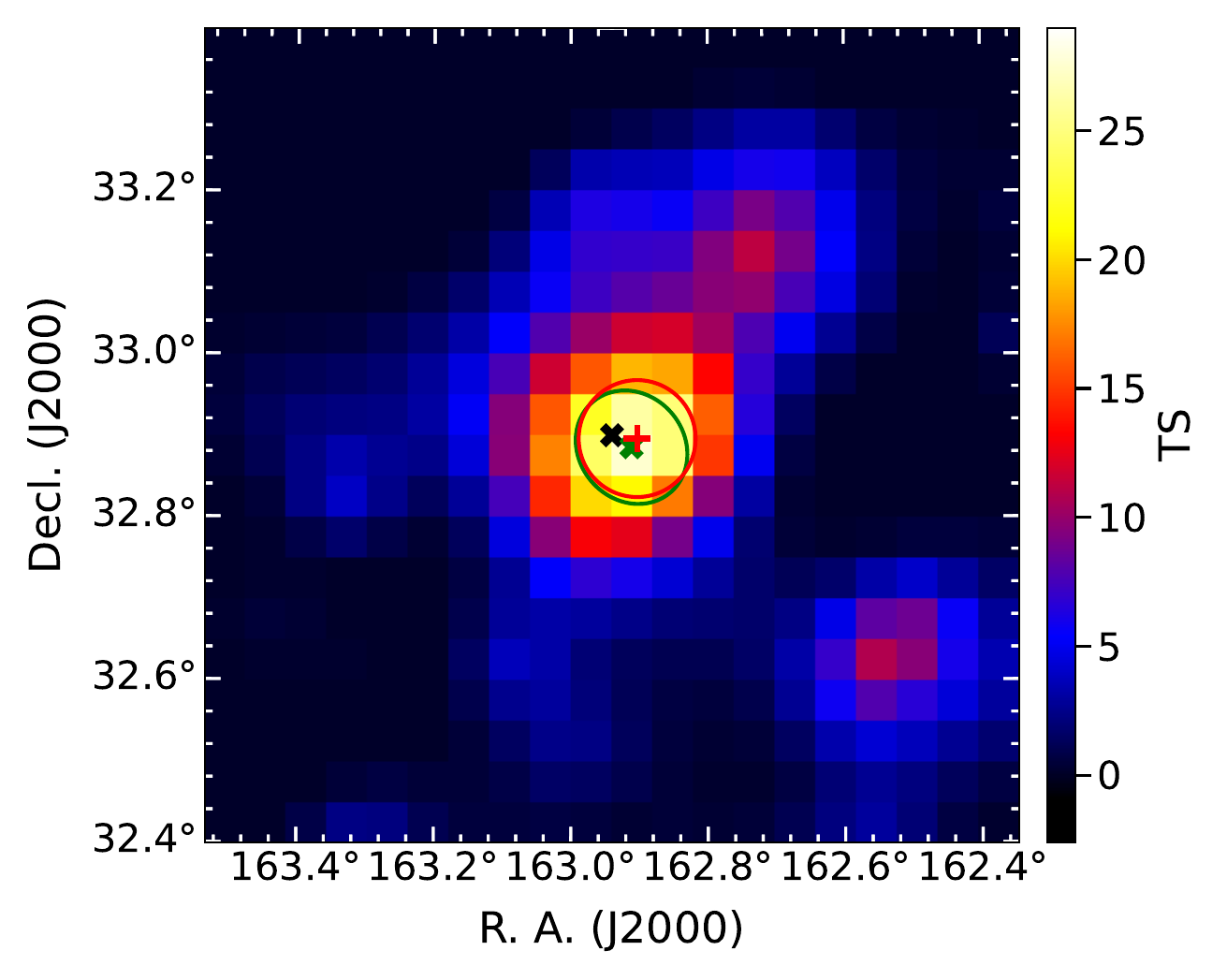}
\includegraphics[width=8.4cm, angle=0]{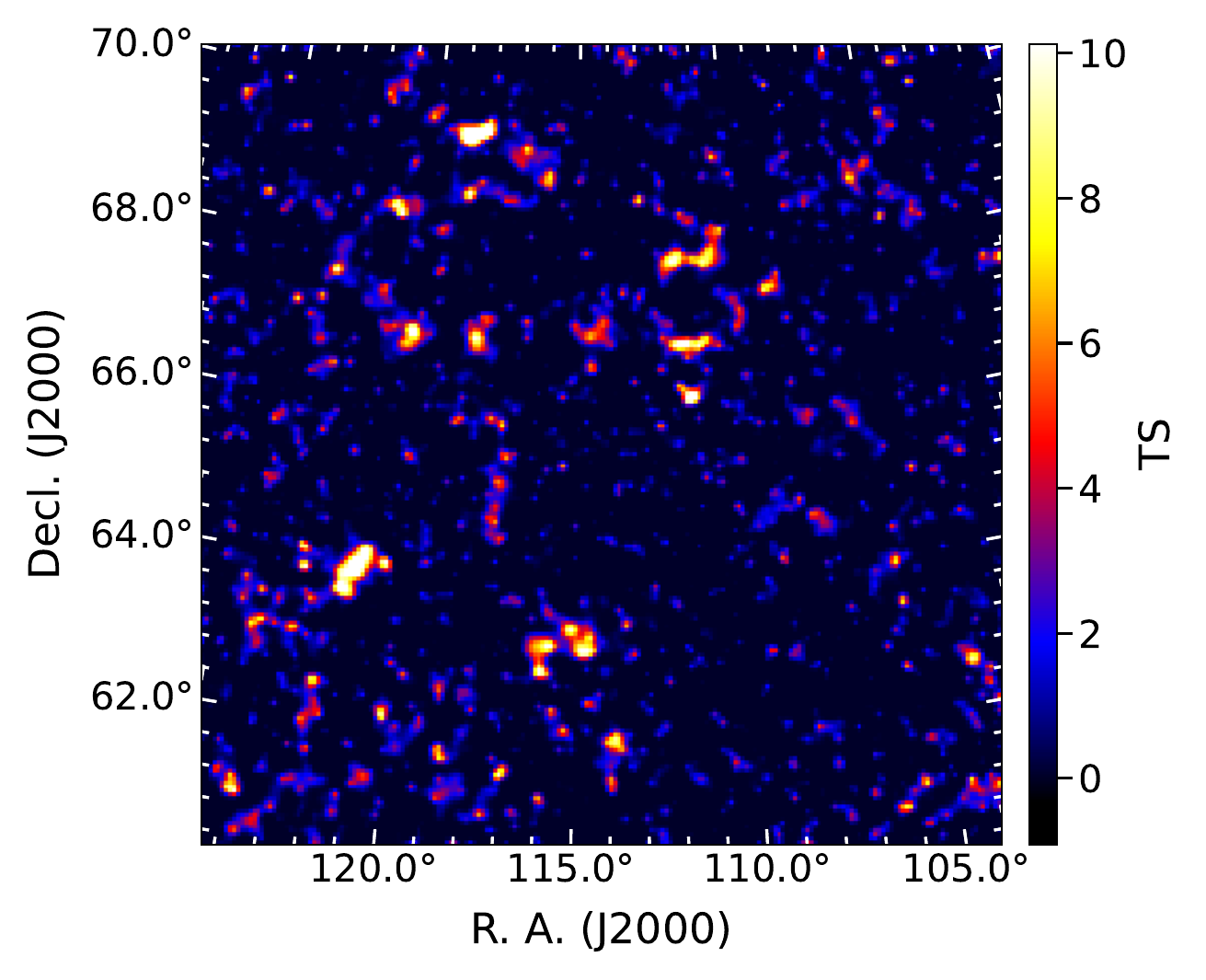}
\includegraphics[width=8.4cm, angle=0]{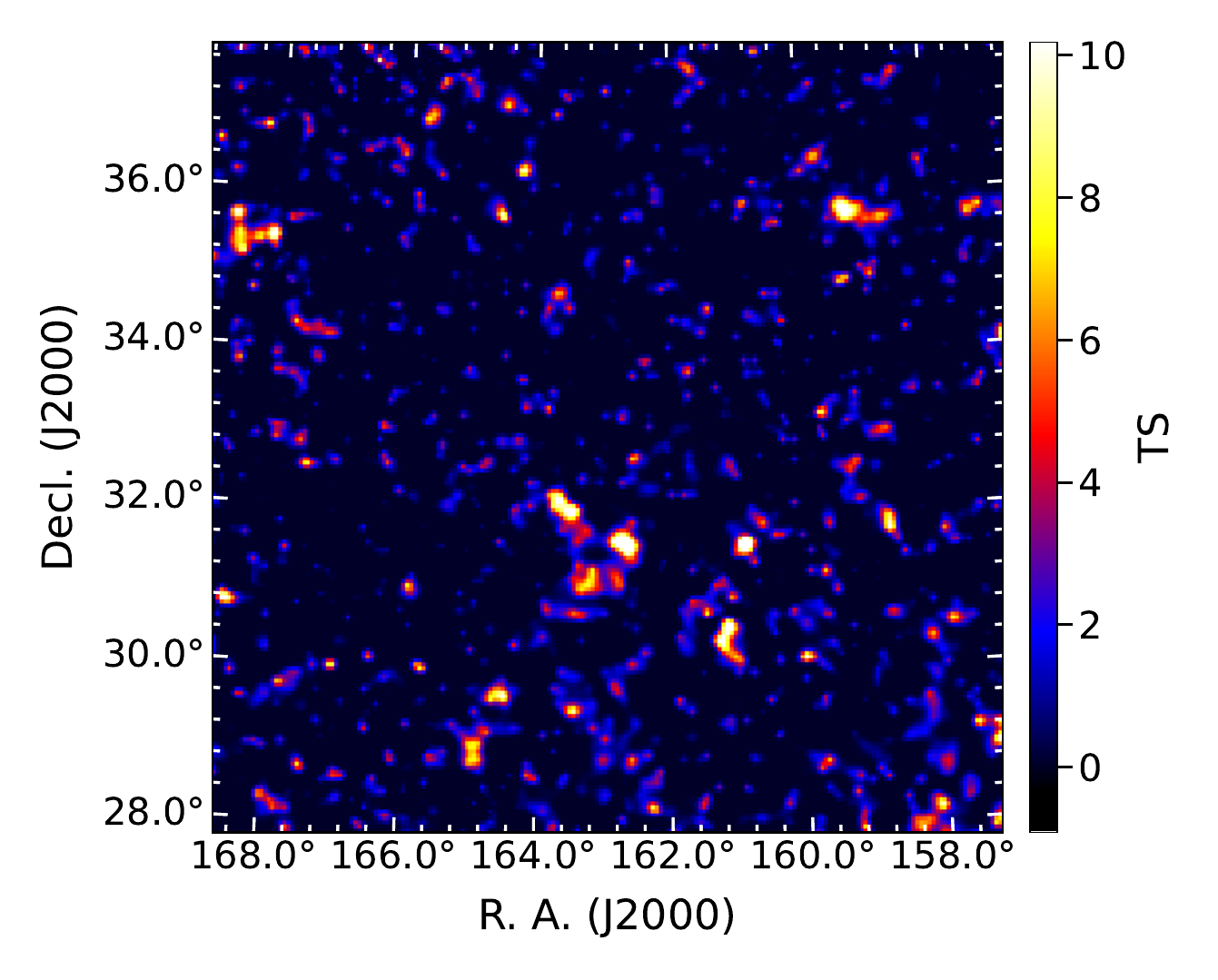}
\caption{Top row: TS maps in the energy band of 1$-$800 GeV covering the $1^{\circ}\times 1^{\circ}$ region around NGC 2403 (left) and NGC 3424 (right). Top left: overlaid are the optical position of NGC 2403 (black cross), the 4FGL J0737.4+6535 position and the corresponding 95\% localization error ellipse (green cross and ellipse), the best-fit point source position (red plus symbol) and the 95\% confidence localization radius (red circle), the optical position of SN 2004dj (blue cross). Top right: overlaid are the IR position of NGC 3424 (black cross), the 4FGL 1051.6+3253 position and the corresponding 95\% localization error ellipse (green cross and ellipse), the best-fit point source position (red plus symbol) and the 95\% confidence localization radius (red circle). Bottom row: TS residual maps for NGC 2403 (left) and NGC 3424 (right) in a $10^{\circ}\times 10^{\circ}$ region. All maps have a pixel size of $0\fdg05$. }
\label{fig:tsmaps}
\end{figure*}

The top panels of Figure ~\ref{fig:tsmaps} show the $1^{\circ}\times 1^{\circ}$ TS maps for NGC 2403 and NGC 3424 after removing the contributions of all sources in the source model except 4FGL J0737.4+6535 (left panel) and 4FGL J1051.6+3253 (right panel), respectively. 
It is clear that SN 2004dj falls in the $r_{95}$ ($0\fdg057$) of our best-fit point source position of the $\gamma$-ray source, while NGC 2403 is completely out of it (offset $0\fdg066$). Our result suggests that the $\gamma$-ray emission is inconsistent with NGC 2403, and it likely originates from supernova SN 2004dj as previously reported \citep{Xi+etal+2020a}. 
To better assess this hypothesis, we tested the MIPS 24 $\mu$m IR template of NGC 2403 $-$ which traces star-forming activity $-$ in two cases: using the template alone and combining it with a point source at the location of SN 2004dj. Both models resulted in a degraded fit quality, thus disfavoring the IR disk template (TS$<4$).
For NGC 3424, the galaxy falls perfectly within the $r_{95}$ of the best-fit point source position, suggesting that the $\gamma$-ray emission is spatially coincident with NGC 3424.  
The bottom panels of Figure~\ref{fig:tsmaps} show the TS residual maps covering a larger region than the TS maps, demonstrating that the modeling of the ROI is of good quality, with no significant residuals (TS$>$16) remaining in the maps. 
In the end, we took the relocalized point source model as the best-fit model for each galaxy and used them in the subsequent spectral and variability analyses, although the improvement over the 4FGL position is not significant.

\subsection{Spectral Analysis}
\label{spectral}
For the spectral analysis of the target galaxies, we performed a binned summed maximum likelihood fitting in the 0.1$-$800 GeV energy range. To test the spectral properties in $\gamma$-rays, we modeled the spectrum of both NGC 2403 and NGC 3424 using two spectral models: a PL model
 \begin{equation}
\frac{dN}{dE}~=~N_0 \left(\frac{E}{E_0} \right) ^{-\Gamma}
\label{eq:powerlaw}
 \end{equation}
 where $N_0$ is the normalization, $E_0$ is the scale parameter that is fixed to 1 GeV, and $\Gamma$ is the spectral index; and a log-parabola (LP) model
 \begin{equation}
\frac{dN}{dE}~=~N_0 \left(\frac{E}{E_b} \right) ^{-(\alpha+\beta \ln(E/E_b))}
\label{equ:lp}
 \end{equation}
where $E_b$ is the scale parameter that is also fixed to 1 GeV, $\alpha$ is the spectral index at $E_b$, and $\beta$ is the curvature index.
For both NGC 2403 and NGC 3424, the LP model did not improve the fit significantly (TS$_{\rm curv}<9$). We adopted PL as the best-fit model in the subsequent analysis.
We derived the photon and energy fluxes above 100 MeV as $F_{\rm 100}$ and $G_{\rm 100}$, respectively. The $\gamma$-ray luminosity of each galaxy was calculated by $L_{\gamma} = 4\pi d^{2} G_{\rm 100}$, with $d$ the distance to the galaxy, which are 3.2 Mpc and 27.2 Mpc for NGC 2403 and NGC 3424, respectively. Table~\ref{tab:Tab3} summarizes the log-likelihood value of the PL model fit, the best-fit spectral parameters, TS, TS$_{\rm curv}$, as well as the derived fluxes and luminosities for both galaxies.

\begin{figure*}[htbp]
\centering
\includegraphics[width=8.5cm, angle=0]{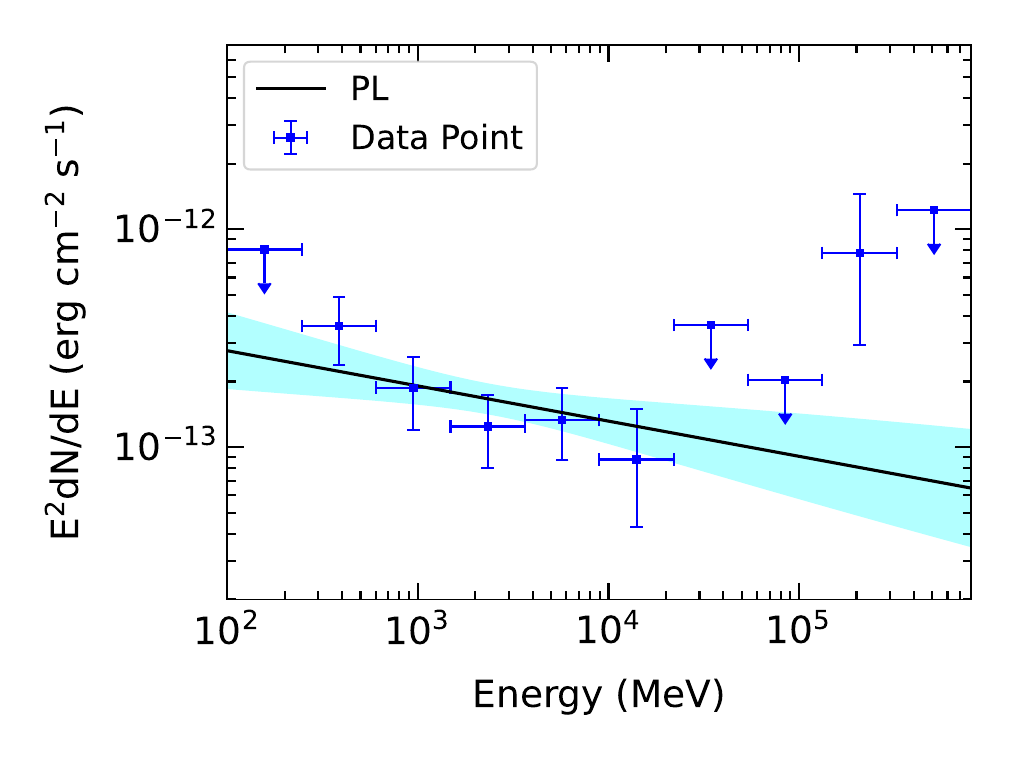}
\includegraphics[width=8.5cm, angle=0]{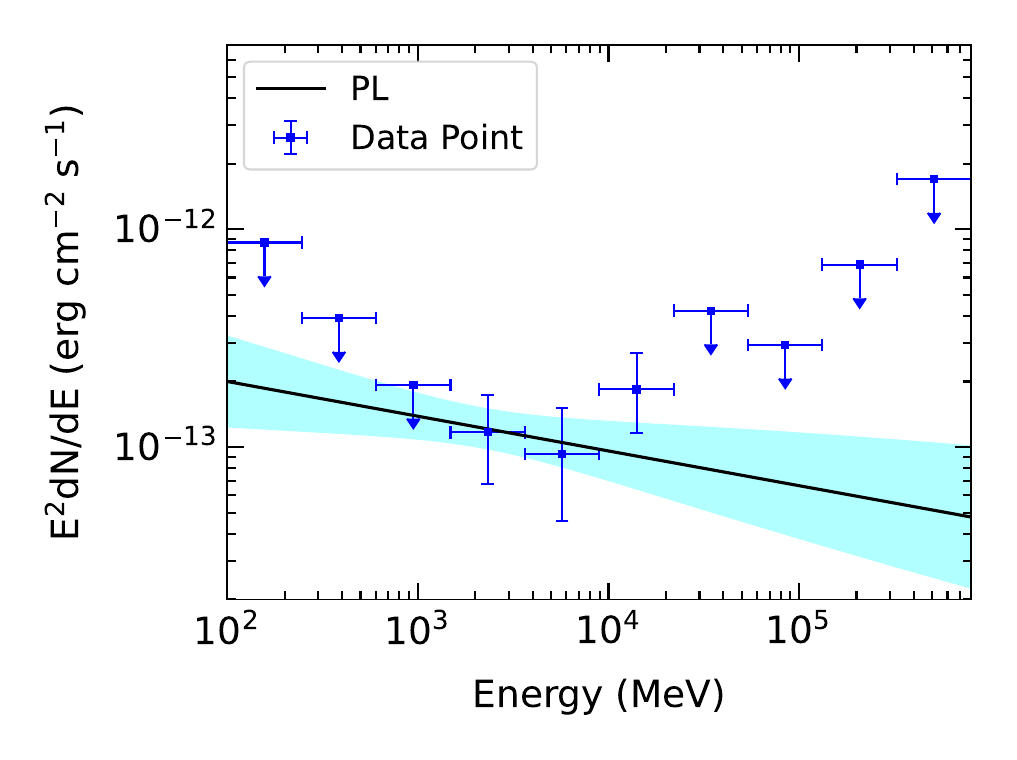}
\caption{SEDs for NGC 2403 (left) and NGC 3424 (right) in the 0.1$-$800 GeV energy range. The black solid line represents the best-fit PL model with $1\sigma$ confidence level uncertainty (light cyan shaded region) from the broadband analysis. The blue data points show the energy flux in each time bin, and the blue arrows are upper limits at the $2\sigma$ confidence level, obtained in the energy bins in which the source has a TS less than 4.} 
\label{fig:SEDs}
\end{figure*}

We generated the spectral energy distribution (SED) by performing a maximum likelihood analysis in 10 logarithmic spaced energy bins over 0.1$-$800 GeV. Within each energy bin, we allowed the normalization parameters of all background components to vary.
The SEDs of NGC 2403 and NGC 3424 are shown in Figure ~\ref{fig:SEDs}, together with the best-fit PL models. The flux upper limit at the 95\% confidence level was calculated when the source has a TS $<$ 4 in a given energy bin.

\begin{deluxetable*}{lcccccccc}
\tabletypesize{\footnotesize}
\tablewidth{\textwidth} 
\tablecaption{Spectral Analysis Results in 0.1--800 GeV for NGC 2403 and NGC 3424}
\label{tab:Tab3}
\tablehead{
\colhead{Model} & \colhead{$-\log \mathcal{L}$} & \colhead{TS} & \colhead{TS$_{\rm curv}$} & \colhead{$\Gamma$} & \colhead{$F_{\rm 100}$} &
\colhead{$G_{\rm 100}$} & \colhead{$L_{\rm \gamma}$} & \colhead{$\log_{10}(L_{\gamma})$} \\
\colhead{(Spatial--Spectral)} & & & & & \colhead{($10^{-9}$ ph cm$^{-2}$ s$^{-1}$)} & \colhead{($10^{-12}$ erg cm$^{-2}$ s$^{-1}$)} & \colhead{($10^{39}$ erg s$^{-1}$)} &
}
\startdata
\multicolumn{9}{c}{NGC 2403} \\
\hline
Point source + PL & 1370680.5 & 52.6 & 2.2 & $2.16 \pm 0.16$ & $1.49 \pm 0.62$ & $1.31 \pm 0.24$ & $1.61 \pm 0.30$ & 39.21 \\
\hline
\multicolumn{9}{c}{NGC 3424} \\
\hline
Point source + PL & 892008.8 & 30.3 & 0.0 & $2.16 \pm 0.19$ & $1.07 \pm 0.52$ & $0.96 \pm 0.24$ & $84.54 \pm 21.37$ & 40.93 \\
\enddata
\tablecomments{
The quantity $-\log \mathcal{L}$ is the negative log-likelihood value of the model fit.
TS and TS$_{\rm curv}$ denote the detection significance and the spectral curvature
significance, respectively. $\Gamma$ is the photon index of the PL model.
$F_{\rm 100}$ and $G_{\rm 100}$ are the photon and energy fluxes derived from the PL fit.
$L_{\rm \gamma}$ is the $\gamma$-ray luminosity assuming distances of 3.2~Mpc and
27.2~Mpc for NGC~2403 and NGC~3424, respectively. Only statistical uncertainties are shown.
}
\end{deluxetable*}

\subsection{Flux Variability}
\label{variability}

To investigate the variability of the $\gamma$-ray flux in the direction of NGC 2403, we generated light curves with 1 yr binning (17 time bins) and 17.1 month binning (12 time bins) over more than 16.5 yr in the energy range of 0.1$-$800 GeV. The latter binning was adopted to be consistent with \cite{Xi+etal+2020a}.
For NGC 3424, the light-curve binning is 1 yr and 8.4 month (24 time bins), with the latter following \cite{Peng+etal+2019}.
For each time bin, we performed an independent maximum likelihood analysis using the best-fit PL model obtained previously for each galaxy.
It is worth mentioning that we have eliminated the impact of solar flares
and GRBs by subtracting them from the dataset. The upper limit at the 95\% confidence level was calculated when the $\gamma$-ray source had TS $<$ 4 in a given time bin. 

To quantify the variability of the light curves, we followed the method presented in \citet{4FGL} by computing the variability index TS$_{\rm var}$, defined as
\begin{equation}
\begin{split}
\rm{TS}_{\rm var} &= 2\textstyle \sum_{i}\!\left[\frac{\log\mathcal{L}_i(F_i)} {\log\mathcal{L}_i(F_{\rm glob})}\right] -\max(\chi^2(F_{\rm glob})-\chi^2(F_{\rm av}), 0)
\end{split}
\label{equ:TS_var}
\end{equation}

\begin{equation}
\chi^{2}(F)=\sum_{i}\frac{(F_{i}-F)^2}{\sigma_i^{2}},
\label{equ:chi2}
\end{equation}
where $F_i$ and $\sigma_i$ are the flux and its statistical uncertainty in the $i$th time bin; $F_{\rm glob}$ is the best-fit flux obtained from the broadband analysis; and $F_{\rm av}$ is the average flux from the light curve. 
The first term in Equation (3) is identical to Equation~(4) in the 2FGL paper \citep{Nolan+etal+2012}. The second term provides a $\chi^2$-based correction, subtracted only when positive, to account for the difference between $F_{\rm glob}$ and $F_{\rm av}$. In our calculation, for bins with TS$<1$ and very small errors, we replaced $\sigma_i$ by the upper error computed following Section 3.6 of the 2FGL paper.

\begin{figure*}[htbp]
\centering
\includegraphics[width=0.45\linewidth]{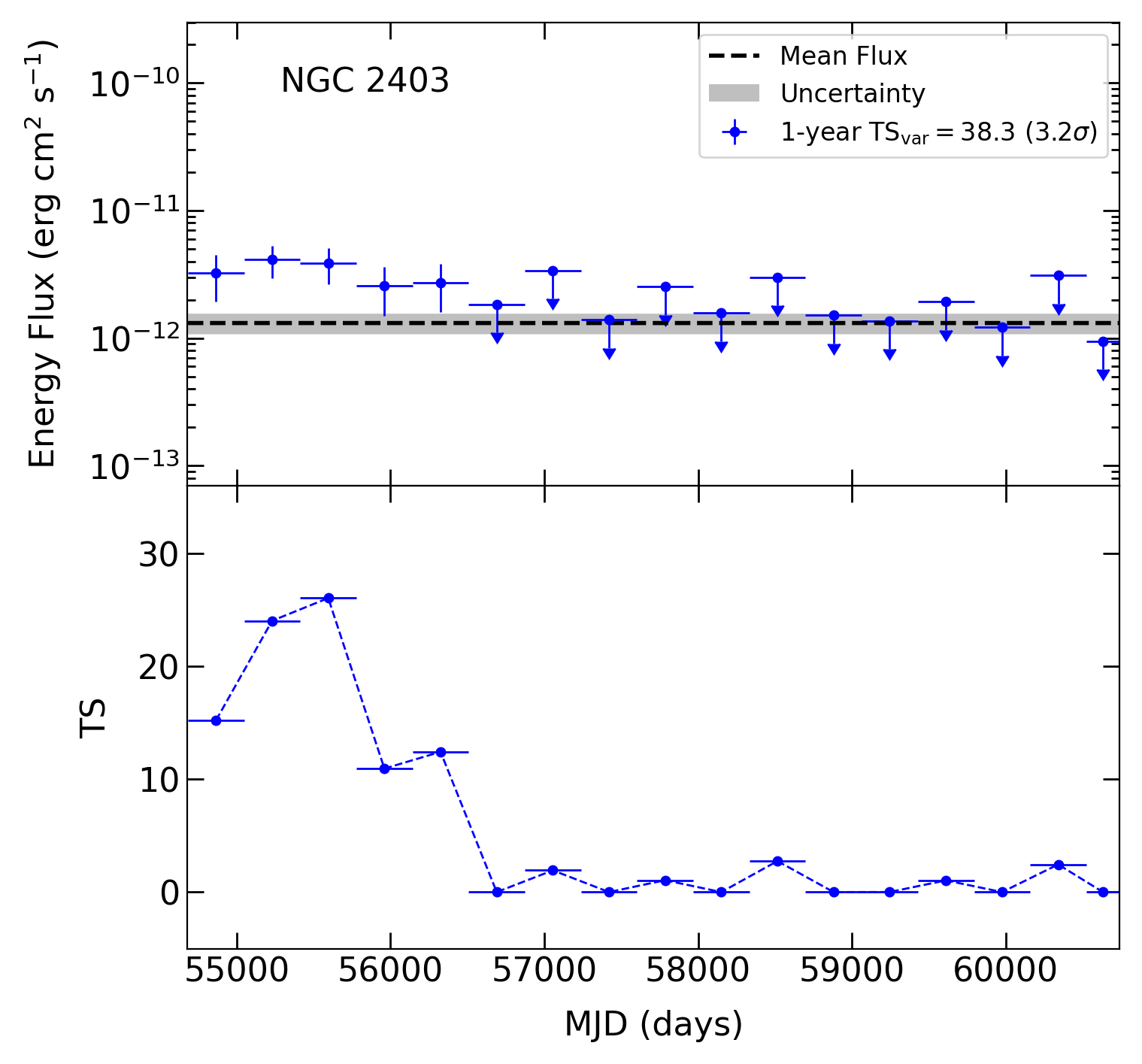}
\includegraphics[width=0.45\linewidth]{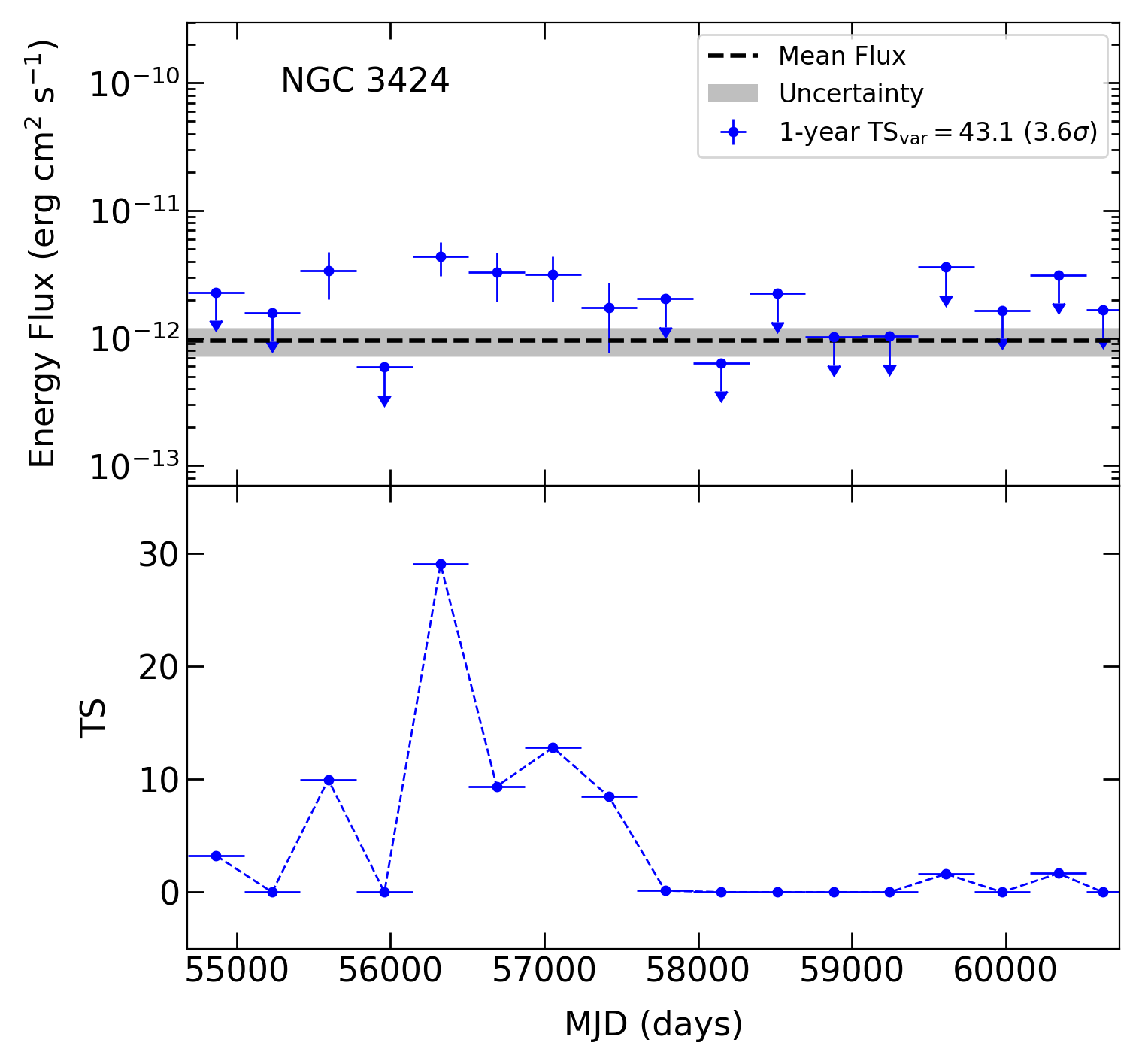}
\caption{
Light curves and TS evolution in the 0.1$-$800 GeV range with 1 yr time binning for NGC 2403 (left) and NGC 3424 (right).
The black dashed line and gray-shaded region represent the best-fit flux and the 95\% uncertainty obtained from the broadband analysis.  
Arrows indicate the 95\% confidence upper limits obtained when the source TS is less than 4 in a given bin. }
\label{fig:lightcurves_NGC2403_NGC3424}
\end{figure*}

For NGC 2403, we obtained TS$_{\rm var}=38.3$ ($\sim3.2\sigma$, with 16 degrees of freedom) with 1 yr binning, and 35.0 ($\sim3.7\sigma$, with 11 degrees of freedom) with 17.1 month binning\footnote{In a $\chi^{2}$ distribution with 16 degrees of freedom, variability is considered probable when TS$_{\rm var}$ exceeds the threshold of 32.00, corresponding to a 99\% confidence level. For distributions with 11 and 23 degrees of freedom, the respective threshold values are 24.72 and 41.64.}, suggesting significant flux variation of the source.
The 1 yr binning light-curve variability is consistent with the 4FGL-DR4 result (TS$_{\rm var}=35.6$, 3.4$\sigma$), and the 17.1 month binning light curve is roughly consistent with \cite{Xi+etal+2020a}, which reported a 2.7$\sigma$ level variability based on the 11.4 yr data analysis. The 1 yr binning light curve is shown in the left panel of Figure~\ref{fig:lightcurves_NGC2403_NGC3424}.
The light curve shows a clear increase in flux and TS, followed by a gradual decline until the source became undetectable, indicating the transient nature of its $\gamma$-ray emission.

For NGC 3424, we obtained TS$_{\rm var}=43.1$ ($\sim3.6\sigma$, with 16 degrees of freedom) with 1 yr binning, and 46.3 ($\sim3.0\sigma$, with 23 degrees of freedom) with 8.4 month binning, suggesting also significant flux variations for NGC 3424.
The 1 yr binning light-curve variability is consistent with the 4FGL-DR4 result (TS$_{\rm var}=41.7$, 4.0$\sigma$), and the 8.4-month binning variability is marginally consistent with \cite{Peng+etal+2019}, which reported significant (TS$_{\rm var}=45.3$, 4.1$\sigma$) variability in the $\gamma$-ray emission of NGC 3424 over 10.5 yr of data. 
The 1 yr binning light curve is shown in the right panel of Figure~\ref{fig:lightcurves_NGC2403_NGC3424}. Around MJD 56500, the emission from NGC 3424 reaches a maximum, marked by a prominent flare that persists for roughly 200 days, indicative of substantial activity on a yearly timescale.

\section{Discussion}
\label{discussion}

\subsection{NGC 2403}

The spatial analysis shows that NGC 2403 is significantly offset from the best-fit $\gamma$-ray emission position (offset $0\fdg066$, outside of the $r_{95}$ of $0\fdg057$). The optical size (radius) of NGC 2403 is $0\fdg167$ in radius \citep{Makarov2014}, while its IR size is only $0\fdg051$ \citep{Jarrett2003}, which is also smaller than the $r$95 of the best-fit $\gamma$-ray source. This would disfavor the association of the observed $\gamma$-ray source with NGC 2403. 
Furthermore, the observed $\gamma$-ray emission is fading with time, as previously reported in \cite{Xi+etal+2020a}. Such significant long-term flux variability contradicts the steady emission expected from the CR-ISM interaction process in SFGs \citep{Thompson2007}. 
Finally, the $\gamma$‑ray luminosity we derived deviates from the canonical $L_{\rm \gamma}$-$L_{\rm IR}$ relation (log$_{10}L_{\gamma}\sim$ 39.21), lying roughly a factor of 10 above the correlation reported in \cite{Ajello+etal+2020} and well exceeding the calorimetric limit. This result is consistent with earlier studies and provides additional evidence against a CR-ISM origin for the emission, since the correlation primarily relies on the CR–ISM interaction.

Our best-fit position of the $\gamma$-ray source is consistent within uncertainties with that reported in \cite{Xi+etal+2020a} who used the first 5.7 yr of data for the localization given the non-detection of the source in the second 5.7 yr inferred from the long-term light curve of the source (also visible from our Figure \ref{fig:lightcurves_NGC2403_NGC3424}). Based on the NASA/IPAC Extragalactic Database (NED) and SIMBAD, there are no Galactic or extragalactic $\gamma$-ray sources within the $r_{95}$ of the $\gamma$-ray source. 
\cite{Xi+etal+2020a} calculated the expected $\gamma$-ray fluxes of two radio sources with unknown nature (NVSS J073724+653628 and [ECB2002] NGC 2403 alf) falling in their $r_{95}$ and found that they are two orders of magnitude lower than observed, and their multiband nature does not support common high-energy interpretations such as blazars or X-ray binaries. Within our $r_{95}$, we found a third radio source, CB6 B0732+6542. Following the method in \cite{Xi+etal+2020a}, the expected $\gamma$-ray flux of this source is also one order of magnitude lower than observed. Thus, these three radio sources are less likely to be the counterpart of the $\gamma$-ray source. 

The Type IIP SN 2004dj is perfectly within the $r_{95}$, and the probability of it overlapping with a $\gamma$-ray source by chance is extremely low \citep[0.22\%;][]{Xi+etal+2020a}, significantly ruling out the possibility of a random background source. \cite{Xi+etal+2020a} proposed a hadronic model to explain the observed $\gamma$-ray emission where the SN ejecta interacts with a dense circumstellar shell produced by the stellar wind of the progenitor star \citep{Dwek1985}, i.e., a red supergiant for this kind of Type IIP SN. 
This interaction drives a shock wave that accelerates protons, which then produce the detected $\gamma$-rays via proton-proton collisions and the subsequent decay of neutral pions. Assuming that the shock was produced shortly after the SN explosion in 2004, say, around the start of the Fermi mission in 2008, this model can explain the observed total $\gamma$-ray energy and the flux decay due to the stopping of proton acceleration and the proton cooling in 5.7 yr. Indeed, the ongoing circumstellar interaction is corroborated by a clear radio detection of SN 2004dj with the Very Large Array at 1.425 GHz on 2008 Oct 24 \citep[$\sim$1578 days after explosion;][]{Nayana2018}. 

To assess the scenario, we estimated the expected $\gamma$-ray emission following \citet{Dwarkadas2013} and \citet{Cristofari2020}. 
For an SN expanding into a circumstellar medium (CSM), the intrinsic $\gamma$-ray luminosity scales as
\begin{equation}
L_\gamma(>E_0,t) \propto \left(\frac{\dot{M}}{v_w}\right)^2 t^{m-2} \, ,
\end{equation}
where $E_0$ is the lower energy threshold for the integrated luminosity (typically 100 MeV), $\dot{M}$ is the mass-loss rate of the progenitor star, $v_w$ is the wind velocity, and $m \approx 0.85–0.9$ is the shock expansion index \citep{Dwarkadas2013}. 
For a Type IIP SN such as SN 2004dj, the progenitor’s steady-state wind is characterized by $\dot{M} \approx1 \times 10^{-6}~M_\odot~\mathrm{yr^{-1}}$ and $v_w = 10~\mathrm{km~s^{-1}}$ \citep{Nayana2018}. Evaluating the model at the epoch of the $\gamma$-ray detection ($t \approx 4$ yr postexplosion) and using the full analytical expression of \cite{Dwarkadas2013} (Eq.9) with standard core-collapse SN parameters result in an expected $L_\gamma \sim 10^{37}~\mathrm{erg\,s^{-1}}$, roughly two orders of magnitude below the observed value of $L_\gamma \approx 1.61 \times10^{39}~\mathrm{erg~s^{-1}}$ (Table~\ref{tab:Tab3}). 
Reconciling this discrepancy requires an enhancement of the CSM density by a factor of $\sim 10$ relative to a standard red supergiant wind, corresponding to an effective mass-loss rate $\dot{M}_{\rm eff} \sim 10^{-5}~M_\odot~\mathrm{yr^{-1}}$. This discrepancy in $\gamma$-ray luminosity weakens the scenario of an association with SN2004dj or suggests that the SN-CSM interaction might be more complex. One possibility is that radio and $\gamma$-ray observations probe different regions in an asymmetric or inhomogeneous CSM environment.

\subsection{NGC 3424}

As revealed by the spatial analysis, NGC 3424 lies well within the $r_{95}$ ($0\fdg072$) of the best‑fit $\gamma$‑ray position, even taking its optical radius \citep[$0\fdg021$,][]{Makarov2014} and IR radius \citep[$0\fdg019$,][]{Jarrett2003} into account. In addition, no other known sources are present within this region, indicating a robust positional association.
Similar to NGC 2403, the long-term light curve of NGC 3424 exhibits significant variability, and its luminosity (log$_{10}L_{\gamma}\sim$ 40.93) deviates from the $L_{\rm \gamma}$-$L_{\rm IR}$ correlation by a factor of approximately 10, in agreement with previous studies \citep{Peng+etal+2019,Ajello+etal+2020}. 
Indeed, NGC 3424 is spatially associated with 4FGL J1051.6+3253 in the 4FGL-DR4 catalog, which is classified as a variable BL Lac candidate \citep{Germani+etal+2021MNRAS}--a subclass of blazars. 
Our analysis confirms the significant flux variability reported both in the catalog and in \cite{Peng+etal+2019}. Such a variability is a characteristic signature of nonthermal emission from relativistic jets, commonly observed in blazars. This observation supports the identification of the $\gamma$-ray source as an AGN (specifically, a BL Lac candidate) associated with NGC 3424.

Consequently, for NGC 3424, the observed variability, combined with the luminosity excess, strongly favors the obscured AGN or relativistic jet (blazar) scenario over the CR-ISM process. In this context, NGC 3424 may represent a system analogous to NGC 4945 \citep{Wojaczy'nski+etal+2017}, Circinus \citep{Guo+etal+2019}, and NGC 1068 \citep{Ajello+etal+2023}--galaxies hosting both an AGN and a nuclear starburst region (\citealp{Wojaczy'nski+etal+2017, Guo+etal+2019,Ajello+etal+2023}).
Furthermore, the ongoing interaction between NGC 3424 and NGC 3430 likely enhances the star formation activity in NGC 3424, potentially triggering a circumnuclear starburst and/or AGN activation. Such phenomena could serve as indirect evidence for the presence of an AGN in NGC 3424, as suggested by \citealp{Gavazzi+etal+2013A&A} and \citealp{Peng+etal+2019}. 
However, distinct from the steady $\gamma$-ray emission typical of Seyfert 2 galaxies like NGC 1068, the variability detected in NGC 3424 indicates that the high-energy emission is dominated by relativistic jet activity. This observation confirms its association with the variable BL Lac candidate, as only a compact central engine can drive the observed flux changes and high radiative efficiency.

Yet, this interpretation stands in tension with NGC 3424’s classification as a radio-quiet AGN \citep{Kellermann+etal+1989AJ,Gavazzi+etal+2011A&A,Gavazzi+etal+2013A&A,Peng+etal+2019}, which shows no evidence of a relativistic jet, the defining feature of blazars.
To investigate this paradox and search for additional evidence of AGN activity, we conducted a search for X-ray counterparts in SIMBAD within a 4.31$'$ (2$\sigma$) radius around the best-fit position of the $\gamma$-ray emission. This yielded 13 potential counterparts, including two quasars, NGC 3424 and 4FGL J1051.6+3253. None of them has published X-ray spectra. We also searched in the Master X-Ray Catalog\footnote{\url{https://heasarc.gsfc.nasa.gov/w3browse/all/xray.html}} and identified 29 X-ray sources within the same search radius as for SIMBAD. However, detailed information on these sources remains limited, and no definitive X-ray signature confirms the jet activity of NGC 3424. 

The lack of X-ray data for NGC 3424 leaves the obscuration scenario$-$namely, that it harbors a Compton‑thick AGN$-$unverified. For instance, key Compton‑thick tracers are absent: there is no reported strong iron K$\alpha$ line complex in the 6.4$–$7 keV range, nor the characteristic reflection spectrum \citep{Comastri+2004}. 
Future observations$-$such as high-resolution radio VLBI imaging to prove the presence of faint jets and deep X‑ray spectroscopy to constrain the hydrogen column density ($N_{\rm H}$) and to detect any iron K$\alpha$ line$-$are essential for fully resolving the nature of NGC 3424, and for ultimately reconciling its radio-quiet classification with the blazar-like $\gamma$-ray activity uncovered in this work as well as in previous studies.

\section{Conclusions}
\label{conclusion}

We analyzed more than 16.5 yr of \emph{Fermi}-LAT data to characterize the $\gamma$-ray emission toward NGC 2403 and NGC 3424. We find that NGC 2403 is not associated with the detected $\gamma$-ray source, in contrast to NGC 3424. The $\gamma$-ray emissions from both galaxies exhibit significant long-term variability, and they remain outliers in the $L_{\rm \gamma}$-$L_{\rm IR}$ correlation, contrary to the behavior expected for typical SFGs. Our results confirm the findings of previous studies and lead us to conclude that the hypothesis that the $\gamma$-ray emission arises from star-formation-driven CR-ISM processes in both galaxies is disfavored. Instead, the data indicate alternative dominant mechanisms: for instance, an SN ejecta–shell interaction in the case of NGC 2403, although this scenario seems to require relatively high circumstellar densities at first order, and emission from an obscured AGN/BL Lac object for NGC 3424. Future multiwavelength observations, in particular, in radio and X-rays, are required to definitively resolve their natures.


\begin{acknowledgments}
We thank M. Di Mauro and M. Ajello for very useful discussions and comments that helped to improve the article.
This work is supported by the National Key R\&D Program of China (2023YFE0101200). X.H acknowledges support from the National Natural Science Foundation of China under grant No. 12373051. 
The Fermi-LAT Collaboration acknowledges generous ongoing support from a number of agencies and institutes that have supported both the development and the operation of the LAT, as well as scientific data analysis. These include the National Aeronautics and Space Administration and the Department of Energy in the United States; the Commissariat \`a l'Energie Atomique and the Centre National de la Recherche Scientifique/Institut National de Physique Nucl\'eaire et de Physique des Particules in France; the Agenzia Spaziale Italiana and the Istituto Nazionale di Fisica Nucleare in Italy; the Ministry of Education, Culture, Sports, Science and Technology (MEXT), High Energy Accelerator Research Organization (KEK), and Japan Aerospace Exploration Agency (JAXA) in Japan; and the K.~A.~Wallenberg Foundation, the Swedish Research Council, and the Swedish National Space Board in Sweden. Additional support for scientific analysis during the operations phase is gratefully acknowledged from the Istituto Nazionale di Astrofisica in Italy and the Centre National d'\'Etudes Spatiales in France. This work performed in part under DOE contract DE-AC02-76SF00515.
\end{acknowledgments}

\bibliography{main}
\bibliographystyle{aasjournal}

\end{document}